\documentclass[a4paper, 12pt]{article}
\usepackage{epsfig}
\usepackage{graphicx}
\usepackage{amsmath}

\author{Juli\'an Candia\\
{\small\it Center for Complex Network Research and Department of Physics,}\\  
{\small\it Northeastern University, Boston, MA 02115, USA}\\
{\small\it INIFTA, CCT La Plata,    
 Universidad Nacional de La Plata,}\\ {\small\it La Plata, Argentina}\\
{\small Email address: jcandia@nd.edu}
}  

\title{Advertising and irreversible opinion spreading in complex social networks}

\begin{document}
\maketitle

\begin{abstract}
Irreversible opinion spreading phenomena are studied on small-world and scale-free networks 
by means of the magnetic Eden model, a nonequilibrium kinetic model for the growth of 
binary mixtures in contact with a thermal bath. In this model, the opinion of an individual 
is affected by those of their acquaintances, but opinion changes 
(analogous to spin flips in an Ising-like model) are not allowed. We focus on the influence of advertising, 
which is represented by external magnetic fields. 
The interplay and competition between temperature and fields lead to order-disorder transitions, which  
are found to also depend on the link density and the topology of the complex network substrate. The 
effects of advertising campaigns with variable duration, as well as the best cost-effective strategies to 
achieve consensus within different scenarios, are also discussed. 
\end{abstract}

{\it Keywords: } Complex networks; sociophysics; kinetic growth models; phase transitions; nonequilibrium processes.
\vspace{0.4 true cm}

\section{Introduction}
Many interdisciplinary fields of science are nowadays benefitting from valuable tools, insight and 
procedures borrowed from long-standing, traditional applications of statistical 
physics \cite{gal82,oli99,wei00,sta06,cas07}.  
In particular, a rich variety of social phenomena, such as social influence and self-organization, cooperation, 
opinion formation and spreading, have recently been studied using models and methods of statistical physics   
(see e.g. \cite{sch00,szn00,gal02,kup02,vil02,sch03,tes04,gal05,gon05,gon06,
lam06,bor07,maz07,gon07,can08a,gal08}).
In some of these investigations, social processes were studied by means of 
Ising-type spin models \cite{gal97,bar00,git00,ale02,sve02,her02,med03,her04}, in which 
spin states denote different opinions or preferences, while the  
coupling constant describes the convincing power between interacting individuals in competition with 
the ``free will'' given by the thermal noise. 

Four decades ago, seminal observations by Milgram \cite{mil67} showed that the mean social 
distance between a pair of randomly selected individuals was astonishingly short, 
typically of a few degrees. Since then, 
the phenomenon of average shortest path-lengths that scale logarithmically 
with the system size, known as {\it small-world effect}, has been found ubiquitously in many 
different real networks, which also generally show highly interconnected local neighborhoods.   

The well-studied classical random graphs, which are networks built by linking nodes at random, 
display the small-world effect but have much lower connectivities than those usually observed in real networks.   
To address this problem, the so-called {\it small-world networks} were proposed few years ago \cite{wat99, wat98} 
as a realization of complex networks having short mean path-lengths 
(and hence showing the small-world effect) as well as large connectivities. 
Starting from a regular lattice, a small-world network is built by randomly adding or 
rewiring a fraction $p$ of the initial number of links. Even a small fraction of added or rewired links 
provides the shortcuts needed to produce the small-world effect, thus displaying a global behavior close to 
that of a random graph, 
while preserving locally the ordered, highly connected structure of a regular lattice. 

Small-world networks, showing the appropriate topological features observed in real social networks, can thus be  
meaningfully used as spin model substrates in order to investigate different social phenomena. 
In this context, irreversible opinion spreading phenomena on small-world \cite{can06,can07} and other 
complex networks \cite{can07b} have been studied by means of the so-called {\it magnetic Eden model} (MEM), 
a nonequilibrium kinetic growth model in which the deposited particles have an intrinsic spin and grow 
in contact with a thermal bath \cite{aus93,can00,can01} (for a review, see Ref.~\cite{can08}). 
The MEM defined on small-world networks present continuous order-disorder phase transitions taking place 
at finite critical temperatures for any value of the rewiring probability $p>0$ \cite{can06}, a phenomenon analogous to 
observations reported previously in the investigation of equilibrium spin systems \cite{bar00,git00}.

According to the growth rules of the MEM, which are given in 
the next Section, the opinion or decision of an individual are affected by those of their 
acquaintances, but opinion changes (analogous to spin flips in the Ising model) do not occur. 
It should be noticed that this behavior contrasts with other models of opinion formation and evolution, 
which generally consider the possibility of changing opinion. As such, the MEM is applicable to scenarios 
were opinion thermalization processes are unlikely to occur. 

Within this context, the aim of this work is to focus on the influence of advertising, 
which is represented by an external, homogeneous magnetic field. 
The interplay and competition between temperature and fields lead to order-disorder transitions, which 
are investigated by means of order-parameter probability distributions and ensemble averages. The  
corresponding temperature vs. magnetic field order-disorder phase diagram is obtained, and the role played by shortcuts 
is discussed. Moreover, the effects of advertising campaigns with variable duration, as well as the best 
cost-effective strategies to achieve consensus within different scenarios, are also investigated. 

Furthermore, in order to explore effects arising from the nature of the complex network substrate, 
we examine the MEM growing on {\it scale-free networks}. The degree distribution of these networks, $P(k)$, 
which measures the probability that any randomly chosen node has $k$ links, follows a power-law, i.e. $P(k)\sim k^{-\gamma}$. 
Unlike the Poisson degree distribution of classical random graphs and small-world networks, power-law distributions lack any 
characteristic scales. The structure of scale-free networks provides an interesting setting in which social roles 
differ significantly among individuals. Indeed, the most connected nodes in the scale-free 
network (usually designated as {\it hubs}) represent highly influential social leaders, 
whose opinions can potentially affect the decisions of many other individuals in the society.  

This paper is organized as follows: 
in Section 2, details on the model definition and the simulation method are given; 
Section 3 is devoted to the presentation and discussion of the results, and Section 4 contains the conclusions. 

\section{Model and simulation method}
In this work, we consider two paradigmatic types of complex network model: 
(a) the one-dimensional, nearest-neighbor, adding-type 
small-world network (SWN) \cite{bol88,new99}, and (b) the Barab\'asi-Albert (BA) scale-free 
network (SF) \cite{bar99,alb02,dor03}.

In case (a), starting with a ring of $N$ 
sites and $N$ bonds, a SWN realization is built by adding new links connecting pairs of 
randomly chosen sites. For each bond in the original lattice, a shortcut is added with probability $p$. 
During this process, multiple connections between any pair of sites are avoided, as well as  
connections of a site to itself. Since the original lattice bonds are not rewired, 
the resulting network remains always connected in a single component.  
On average, $pN$ shortcuts are added and the mean coordination number is $\langle z\rangle =2(1+p)$.     

In case (b), SF networks are built by following the preferential attachment growth mechanism: 
starting with a small fully-connected graph of size $m$, a new node with $m$ edges   
is added at every time step and is connected to $m$ different nodes already present 
in the system, where the probability for an already 
existing node to acquire a new link is proportional to its degree. The BA type of scale-free network is 
characterized by power-law degree distributions, i.e. $P(k)\sim k^{-\gamma}$, where the degree exponent is $\gamma=3$. 

Once the network is created, a randomly chosen up or down spin is deposited on 
a random site. Then, growth takes place by adding, one by one, further spins to the immediate neighborhood 
(the perimeter) of the growing cluster, while taking into account the corresponding interaction energies. By analogy
to the Ising model, the energy $E$ of a configuration of spins is given by
\begin{equation}
E = - J \sum_
{\langle ij\rangle} S_iS_j - H \sum_i S_i,
\label{energy}   
\end{equation}
where $S_i= \pm 1$ indicates the orientation of the spin for each occupied site (labeled by the 
subindex $i$), $J>0$ is the ferromagnetic 
coupling constant between nearest-neighbor (NN) spins, $H>0$ is the external field, and  
$\langle ij\rangle$ indicates that the summation is taken over all pairs of occupied NN sites.
As with other spin systems defined on complex networks, the magnetic interaction between any pair 
of spins is only present when a network link (either a lattice bond or a shortcut) connects their sites. 

Setting the Boltzmann constant equal to unity ($k_B \equiv 1$), 
the probability for a new spin to be added to the (already grown) cluster is
defined as proportional to the Boltzmann factor exp$(-\Delta E /T)$, 
where $\Delta E$ is the total energy change involved and $T$ is the absolute temperature of the thermal bath. 
Energy, magnetic field and temperature are measured in units of the NN coupling constant, $J$, throughout. 
At each step, all perimeter sites have to be considered and the probabilities of adding a new (either up 
or down) spin to each site must be evaluated. 
Using the Monte Carlo simulation method, all growth probabilities are first computed and normalized, and then    
the growing site and the orientation of the new spin are both determined by means of a pseudo-random number.
Although the configuration energy of a MEM cluster, given by Eq.(\ref{energy}), resembles the Ising Hamiltonian, it 
should be noticed that the MEM is a nonequilibrium model in which new spins are 
continuously added, while older spins remain frozen and are not 
allowed to flip. 
The growth process naturally stops after the deposition of $N$ particles, 
when the network becomes completely filled. 

Since finite-size effects for the MEM growing on small-world and BA scale-free networks are mild 
(see Refs.~\cite{can06,can07b}  for detailed discussions), in this work we adopt a fixed network size, $N=10^3$, 
throughout. For any given set of defining parameters (i.e. the shortcut-adding  
probability $p$, the BA parameter $m$, the magnetic field $H$ and the temperature $T$), ensemble averages were calculated 
over $10^4$ different (randomly generated) networks, and considering 
typically $50$ different (randomly chosen) seeds for each network configuration. 

\section{Results and discussion}

\subsection{Advertising and opinion spreading in small-world networks}
\begin{figure}[t!]
\center \includegraphics[width=5.truein, height=4.1truein]{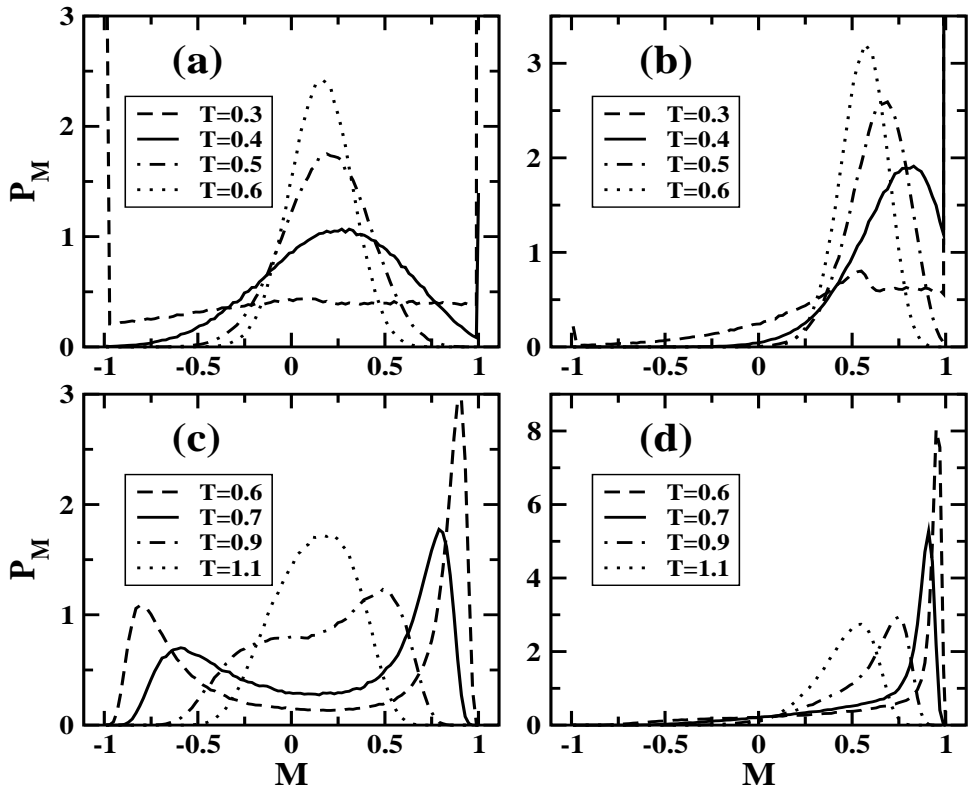}
\caption{Normalized probability distributions of the magnetization for different temperatures, as indicated. 
The panels correspond to different shortcut densities and magnetic fields: 
(a) $p=0$ and $H=0.05$; (b) $p=0$ and $H=0.2$; (c) $p=0.5$ and $H=0.05$; (d) $p=0.5$ and $H=0.2$. 
Sharp maxima at $M=\pm 1$ have been truncated in panels (a)-(b).} 
\label{fig1}
\end{figure}

The natural order parameter of a magnetic system is the total magnetization per site, 
i.e.~\footnote{Notice that, 
although this work is mainly motivated by social phenomena, a magnetic language is 
adopted throughout. Indeed, physical concepts such as temperature and magnetization, 
spin growth and clustering, ferromagnetic-paramagnetic phase transitions, etc, 
can be meaningfully re-interpreted in sociological/sociophysical contexts.}
\begin{equation}
M={\frac{1}{N}}\sum S_i,
\end{equation}
which, in the context of this work, is to be measured on the completely filled network.
The normalized probability distribution of the magnetization, $P_M$, is shown in Figure 1 for 
different temperatures, shortcut densities and magnetic fields: 
(a) $p=0$ and $H=0.05$; (b) $p=0$ and $H=0.2$; (c) $p=0.5$ and $H=0.05$; (d) $p=0.5$ and $H=0.2$. 

For the regular lattice (Figures 1(a)-(b)), the distributions undergo abrupt changes when passing from 
low to high temperatures. In the case of small fields (Figure 1(a)), the low-temperature distribution 
is dominated by sharp maxima taking place at $M=\pm 1$. Increasing temperature, these peaks retract while 
a local maximum develops near $M\approx 0.2$ forming a Gaussian-like distribution. Notice also that, due to 
the small but positive magnetic field, the distributions are skewed and biased towards positive values of 
the magnetization. For large fields (Figure 1(b)), the distributions are highly asymmetric and show the 
same kind of abrupt transition from a low-temperature sharp maximum at $M=1$ to high-temperature Gaussian-like 
shapes.  

\begin{figure}[t]
\centerline{{\epsfxsize=3.9in \epsfysize=2.6in \epsfbox{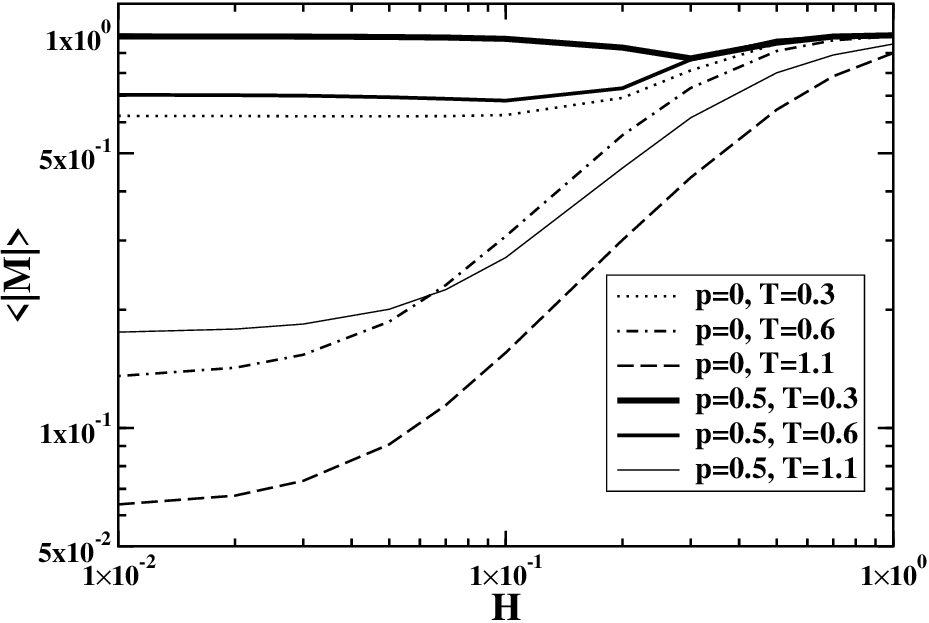}}}
\caption{Ensemble-averaged absolute magnetization as a function of the magnetic field for different 
shortcut densities and temperatures, as indicated.}
\label{fig2}
\end{figure}

In contrast, for the small-world network under small fields (Figure 1(c)) one observes the onset of two
maxima located at $M = -M_{neg}$ and $M = M_{pos}$ (such that $0 < M_{neg} < M_{pos} < 1$),
which become sharper and approach $M= \pm 1$ as $T$ is gradually decreased.
The smooth shift of the distribution maxima across $T\simeq T_c$, from $M\approx 0$ at high $T$ to the low-temperature 
nonzero spontaneous magnetizations $M= -M_{neg}, M_{pos}$, is indeed the signature of true 
thermally-driven continuous phase transitions \cite{can01,bin02}. 
Hence, the presence of shortcuts in the small-world network triggers a critical behavior in the 
irreversible growth of MEM clusters \cite{can06}, analogously to phenomena observed in related equilibrium systems  
such as the Ising model \cite{bar00,git00}. When strong fields are applied (Figure 1(d)), the negative maxima 
at $M = -M_{neg}$ vanish, thus leading to highly asymmetric distributions peaked at $M = M_{pos}$ that tend 
to $M_{pos}\simeq 1$ as $T$ decreases. 

Since, for $H\ll 1$, the low-temperature probability distributions of the magnetization 
are nearly symmetric and sharply peaked around $M\approx\pm 1$, the ensemble-averaged total 
magnetization vanishes. This well-known shortcoming due to finite-size effects is avoided by considering 
the absolute value of the magnetization, $|M|$, as the actual order parameter \cite{bin02,lan00}. 

\begin{figure}[t]
\centerline{{\epsfxsize=3.9in \epsfysize=2.6in \epsfbox{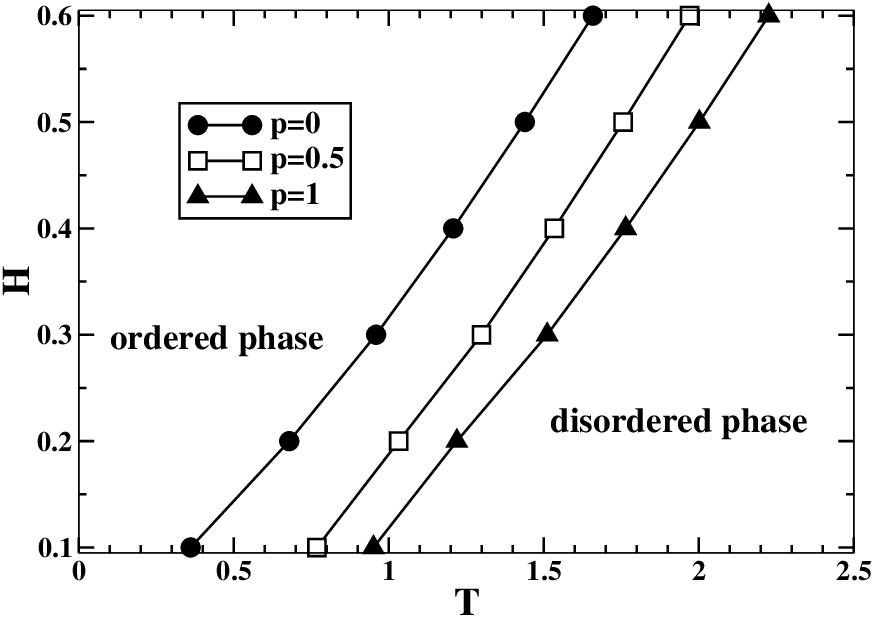}}}
\caption{Phase diagram in the $H$ vs $T$ plane for different values of the shortcut fraction $p$, 
as indicated.}
\label{fig3}
\end{figure}

Figure 2 shows the ensemble-averaged absolute magnetization as a function of the applied magnetic field for 
different shortcut densities and temperatures, as indicated.
For any value of the temperature, the order parameter plot for the small-world network (solid lines) lies above 
the corresponding one for the lattice (dotted/dashed lines), since shortcuts favor long-range ordering 
connections between distant clusters across the network. Moreover, as expected, 
most of the plots are monotonically increasing due to the ordering effect of the external field. 
Interestingly, however, a slight decrease in the absolute magnetization is observed at intermediate fields, 
$H=0.1-0.3$, for $p=0.5$ and $T=0.3$. This behavior is indeed due to the competition between the strong clustering 
tendency of neighboring spins at low temperatures, which favors the formation of extended parallel-oriented 
up and down domains, and the external magnetic field.   

\begin{figure}[t]
\centerline{{\epsfxsize=3.9in \epsfysize=2.6in \epsfbox{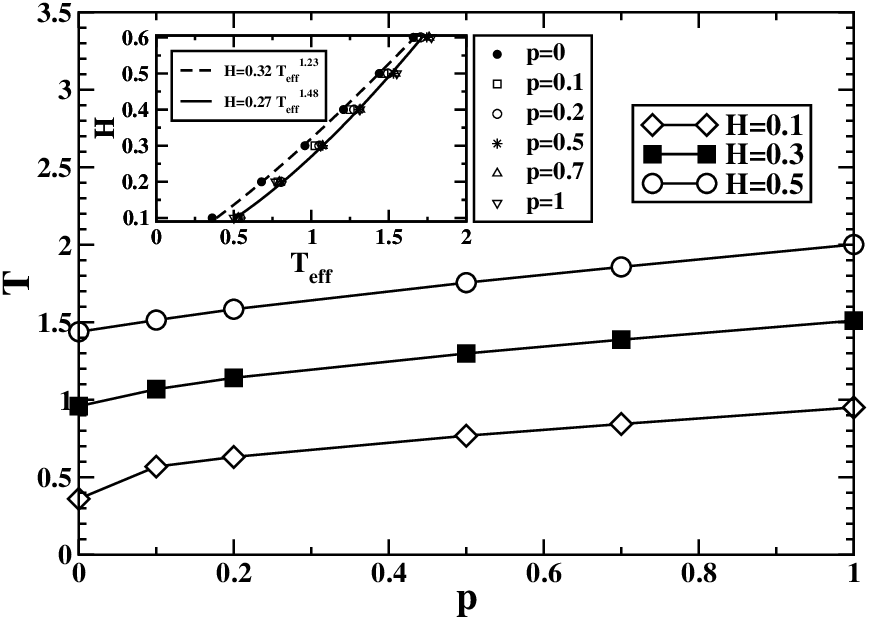}}}
\caption{Order-disorder transition temperature as a function of the shortcut fraction $p$ for different values of 
the external field, as indicated. The inset shows the phase diagram $H$ vs $T_{eff}$, where the effective 
temperature $T_{eff}= T-0.45\times p$ leads to the collapse of the small-world network data. Separate fits 
to the lattice (dashed line) and small-world network (solid line) data are also shown.}
\label{fig4}
\end{figure}

Figure 3 shows $p-$dependent, order-disorder transition curves on a $H$ vs $T$ phase diagram, where 
transition states are defined by the condition $\langle |M|\rangle=0.5$. Due to the long-range 
ordering effect of shortcut connections, transitions for small-world networks with higher shortcut densities 
are shifted towards higher temperatures. Moreover, due to the global reach of the external field, the ordered 
phase is attainable even for very high temperatures. According to the sociophysical interpretation of this model, 
we conclude that consensus can be reached irrespective of the degree of intrinsic cohesion and interconnectedness (as 
given by the parameters $T$ and $p$, respectively) as long as a sufficiently intense, global 
mass media advertising campaign is carried out.  

\begin{figure}[t!]
\center \includegraphics[width=5.truein, height=4.1truein]{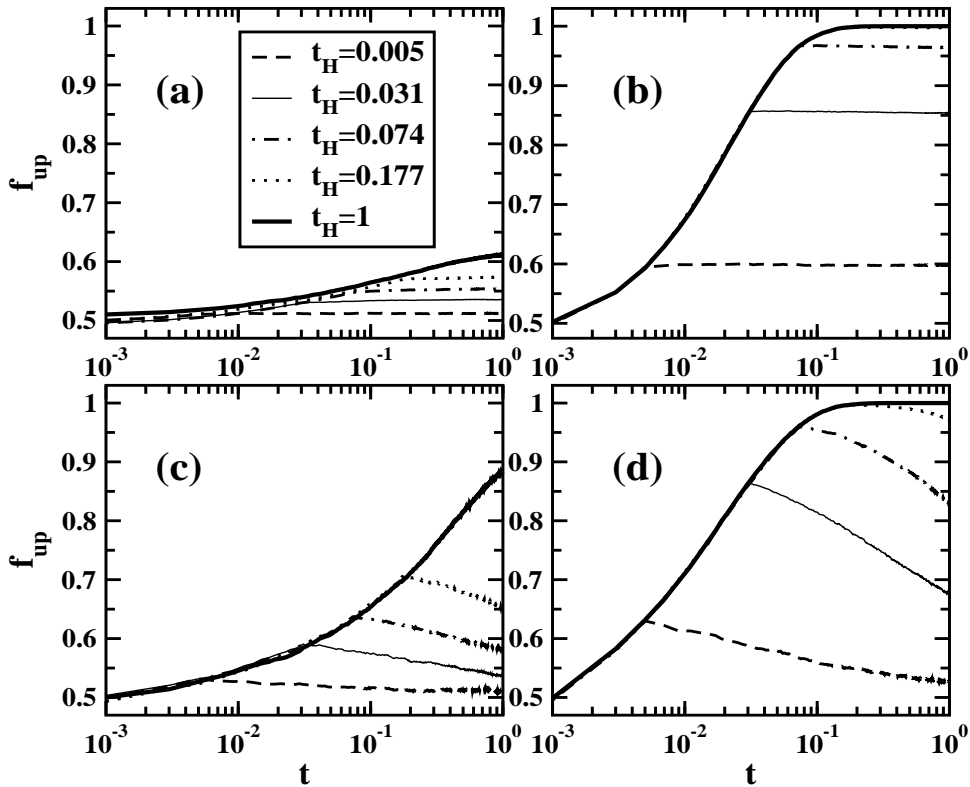}
\caption{Time profiles for the fraction of states with a majority of up spins, $f_{up}$, for small-world networks 
with $p=0.5$ and different field switch-off times, $t_H$, as indicated. 
The panels correspond to different temperatures and magnetic fields: 
(a) $T=0.5$ and $H=0.1$; (b) $T=0.5$ and $H=0.5$; (c) $T=2$ and $H=0.1$; (d) $T=2$ and $H=0.5$.} 
\label{fig5}
\end{figure}

Figure 4 shows the transition temperature as a function of the shortcut fraction $p$ for different values of 
the external field, as indicated. The monotonic shift towards higher temperatures for small-world networks with 
higher connectivities is observed to be roughly independent of the external field. In order to exploit this 
approximate scaling property, we define an effective, shortcut-induced temperature shift $\Delta T_{eff}(p)=-\alpha p$ 
and re-plot the phase diagram in terms of $H$ vs $T_{eff}\equiv T+\Delta T_{eff}$, which is shown in the inset to Figure 4. 
Using $\alpha=0.45$, the small-world network data tend to collapse into a curve that can be well fitted by the 
relation $H=0.27\times T_{eff}^{1.48}$ (solid line). 
As expected, lattice data show a distinct behavior: the $p=0$ transition is well fitted by 
$H=0.32\times T_{eff}^{1.23}$ (dashed line), which clearly departs from the small-world network data collapse. 

\begin{figure}[t!]
\center \includegraphics[width=3.5truein, height=4.7truein]{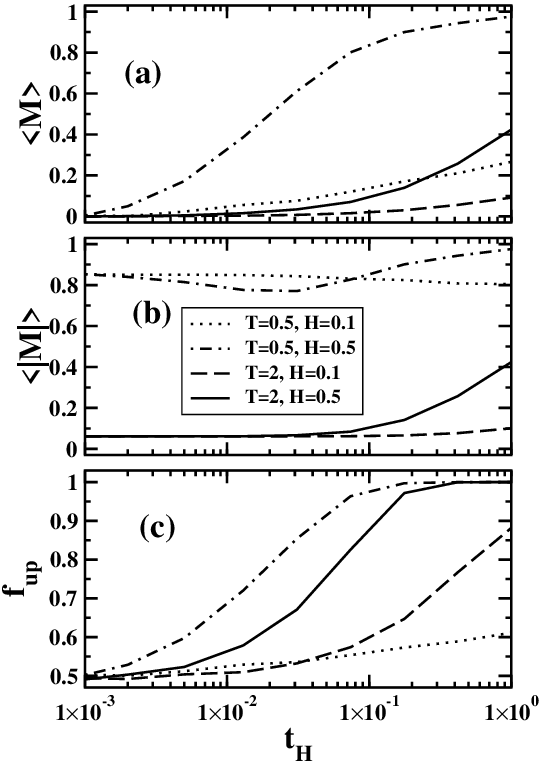}
\caption{Dependence of the ensemble-averaged final states ($t=1$) on the field switch-off time, $t_H$, for different 
temperatures, external fields and observables: (a) the total magnetization, 
(b) the absolute magnetization, and (c) the fraction of final states with a majority of up spins. These results are 
for small-world networks with $p=0.5$.}
\label{fig6}
\end{figure}
 
In order to gain further insight into the dynamical features of the opinion spreading process, let us 
consider the effects of advertising campaigns with variable duration. For the sake of simplicity, we 
assume a constant deposition (or spreading) rate and measure time as $t\equiv n/N$, i.e. the fraction of already deposited 
particles, $n$, relative to $N$, the system's final size. At every stage during the growth process, 
we determine whether a given configuration 
has a majority of up spins ($n_{up} > n/2$) and, after ensemble-averaging over many configurations, we measure the fraction 
of states with a majority of up spins, $f_{up}$. Since the seeds are chosen at random, the initial condition is $f_{up}=0.5$ 
and the success of the campaign can be measured by the extent to which $f_{up}$ approaches unity. We assume that the field 
is applied from $t=0$ to a specified time $t=t_H$; afterwards, the field is switched off. Hence, the 
parameter $0<t_H\leq 1$ represents the duration of the advertising campaign.   

Figure 5 shows time profiles of $f_{up}$ for small-world networks with $p=0.5$, different temperatures and magnetic 
fields: (a) $T=0.5$ and $H=0.1$; (b) $T=0.5$ and $H=0.5$; (c) $T=2$ and $H=0.1$; (d) $T=2$ and $H=0.5$. Different 
field switch-off times correspond to different plots on each panel, as indicated.   
Figure 5(a) shows that small fields are not capable of affecting the strong clustering tendency typical of low 
temperatures. Indeed, the magnetization probability distributions are sharply peaked at $M_{neg}\approx M_{pos}\approx 1$ 
(similarly to the low-temperature plot in Figure 1(c)) and are little influenced by the small external field even 
in the case of $t_H=1$. For low temperatures and strong fields, instead, $f_{up}$ grows steadily with $t_H$ (Figure 5(b)), 
since strong fields deplete the probability distribution for negative values of the magnetization (recall Figure 1(d)). 
Moreover, $f_{up}$ remains constant after the field is switched off due to the lack of sizable thermal fluctuations. 
Notice that highly successful final results (i.e. $f_{up}\sim 1$ for $t\to 1$) can be achieved 
with intense but very short campaigns ($t_H < 0.1$). 
Figure 5(c) shows that, for high temperatures and small fields, $f_{up}$ strongly depends on the campaign duration: 
$f_{up}$ increases monotonically with $t_H$ and is close to unity for $t_H=1$. Comparing to Figure 5(a), one observes that 
higher temperatures increase the overall response sensitivity of the system under the influence of small fields. The case 
of high temperatures and large fields is shown in Figure 5(d). The strong fields naturally drive the system towards 
large values of $f_{up}$; however, whenever the fields are switched off, the trend towards thermal disorder sets back on 
and $f_{up}$ decreases again. 

\begin{figure}[t]
\centerline{{\epsfxsize=3.9in \epsfysize=2.6in \epsfbox{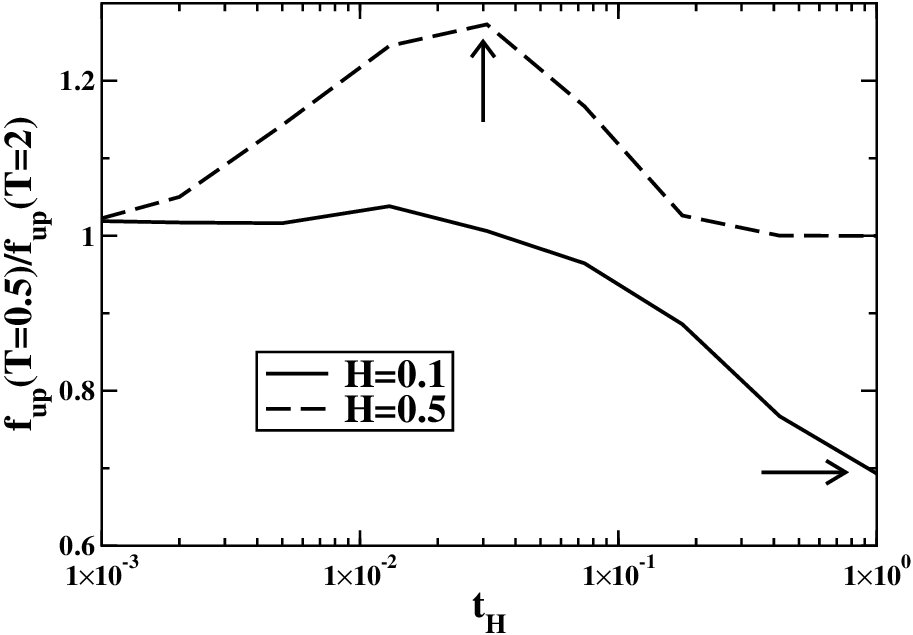}}}
\caption{Ratio $f_{up}(T=0.5)/f_{up}(T=2)$ as a function of the field switch-off time $t_H$ for small-world networks with $p=0.5$ under 
different external fields, as indicated.}
\end{figure}

According to these results, the best cost-effective strategy to achieve a desired majority 
consensus ($f_{up}\to 1$ for $t\to 1$) 
depends crucially on the degree of intrinsic cohesion tendency of the society, here modeled by the temperature. In the case of 
strongly cohesive social groups (low $T$), intense, short campaigns yield larger consensus levels than mild, long ones. 
Interestingly, however, the opposite is true in the scenario of weakly cohesive societies (high $T$)\footnote{Notice that 
the low- and high-temperature regions are relative to the corresponding zero-field critical temperature. 
For SWNs with $N=10^3$ and $p=0.5$, the zero-field transition is located at $T_c\approx 0.6$ \cite{can06}. 
Also recall that temperatures are measured in units of the nearest-neighbor coupling constant throughout 
(see Sect. 2 for details).}.   
 
Figure 6 shows the dependence of the ensemble-averaged final states (i.e. $t=1$) on the campaign 
duration (or, equivalently, 
the field switch-off time, $t_H$) for different temperatures, external fields and observables: (a) the total magnetization, 
(b) the absolute magnetization, and (c) the fraction of final states with a majority of up spins. The results shown 
correspond to small-world networks with shortcut fraction $p=0.5$.  

For low temperatures and strong fields (dash-dotted line plots), the system becomes rapidly ordered in the up direction and 
stays in the up-ordered state even after the field is withdrawn. Hence, $\langle M\rangle$ and $f_{up}$ follow a similar 
pattern, namely a monotonic, steadily increasing behavior as $t_H$ is increased. 
The absolute magnetization is close to unity but displays a dip, which is 
due to the passing of the down-ordered probability distribution, through the zero-magnetization region, towards the 
up-ordered side. This behavior is similar to the dip previously observed in the thick solid line plot of Figure 2.        
For low temperatures and small fields (dotted lines), the positive fields are not capable of significantly affecting the 
strongly clustered down-ordered domains, thus leading to small but moderately increasing values 
for $\langle M\rangle$ and $f_{up}$. In contrast, $\langle |M|\rangle$ is large but slightly decreasing, as explained 
by mild shifting effects of the magnetization probability distribution towards the positive side.   
When high temperatures are considered (long-dashed and solid lines), the magnetization probability distributions 
are nearly Gaussian with positive mean values. Hence, $\langle M\rangle$ and $\langle |M|\rangle$ are expected to 
be similar and generally small. 
However, since in the $H\to 0$ limit these distributions are Gaussians centered at $M=0$, even slight shifts towards $M>0$ 
when small fields are applied lead ultimately to fairly large values of $f_{up}$. This explains the fact that, for high 
temperatures, mild, long campaigns yield higher $f_{up}$ values than short, intense ones.  

Figure 7 shows the ratio $f_{up}(T=0.5)/f_{up}(T=2)$ as a function of $t_H$ for different fields. For $H=0.1$, the ratio 
remains fairly constant about unity up to $t_H\simeq 3\times10^{-2}$, while it decreases monotonically for higher 
values of $t_H$ and reaches a minimum at $t_H=1$ (horizontal arrow), where the 
high-temperature campaign is more successful by more than $40\%$ relative to the low-temperature one. 
Hence, mild, long campaigns are optimal for (high-temperature) weakly cohesive societies. For $H=0.5$, the ratio 
peaks at $t_H=3\times10^{-2}$ (vertical arrow) and then decreases towards unity for larger values of $t_H$, thus showing that for 
(low-temperature) highly cohesive societies, short, intense campaigns are the best strategy.  

\subsection{Advertising and opinion spreading in scale-free networks}

\begin{figure}[t]
\centerline{{\epsfxsize=3.9in \epsfysize=2.6in \epsfbox{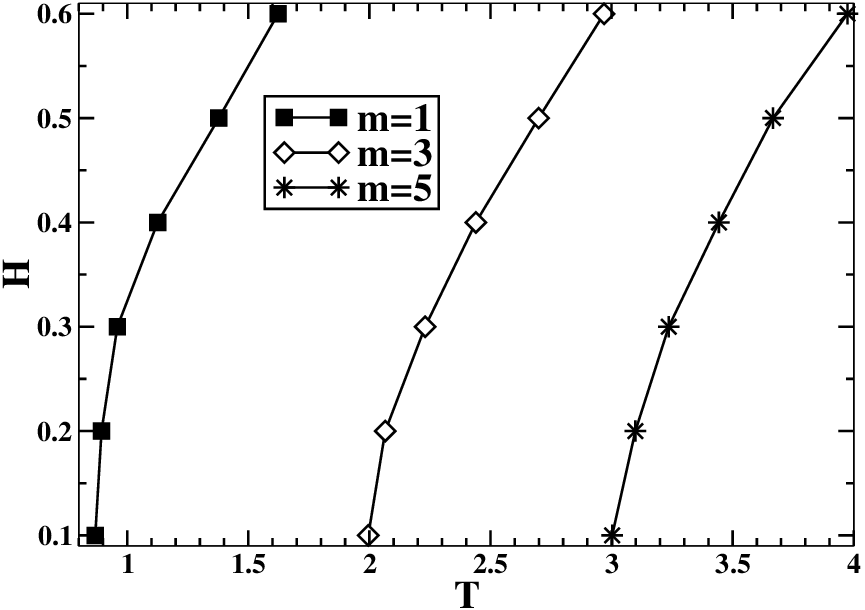}}}
\caption{Phase diagram in the $H$ vs $T$ plane for different values of the Barab\'asi-Albert parameter $m$, 
as indicated.}
\end{figure}

Unlike the case of small-world networks, the highly inhomogeneous degree distribution of 
scale-free networks represents a scenario in which social roles 
differ significantly among individuals. In particular, hubs in the scale-free network 
represent highly influential social leaders, whose opinions can potentially affect the decisions of many other 
individuals in the society.  

The paradigmatic type of scale-free network is based on the preferential attachment growth mechanism: a new node 
with $m$ edges is added at every time step and is connected to $m$ different nodes already present 
in the system, where the probability for an already 
existing node to acquire a new link is proportional to its degree. This simple rule leads to an ensemble of 
Barab\'asi-Albert (BA) scale-free networks, which are characterized by power-law degree distributions with exponent $\gamma=3$. 

Figure 8 shows order-disorder transition plots on a $H$ vs $T$ phase diagram 
for the MEM growing on SF networks with different values of the BA parameter $m$. Since the overall average degree 
grows linearly with $m$, the transition curves for SF networks with higher link densities  
are shifted towards higher temperatures. Moreover, due to the global reach of the external field, the ordered 
phase is attainable even for very high temperatures. Comparing to the SWN results (see Figure 3), we observe that, 
irrespective of the details of the underlying complex network substrate, consensus and order can be reached for any level of 
intrinsic cohesion and interconnectedness.  

\begin{figure}[t]
\centerline{{\epsfxsize=3.9in \epsfysize=2.6in \epsfbox{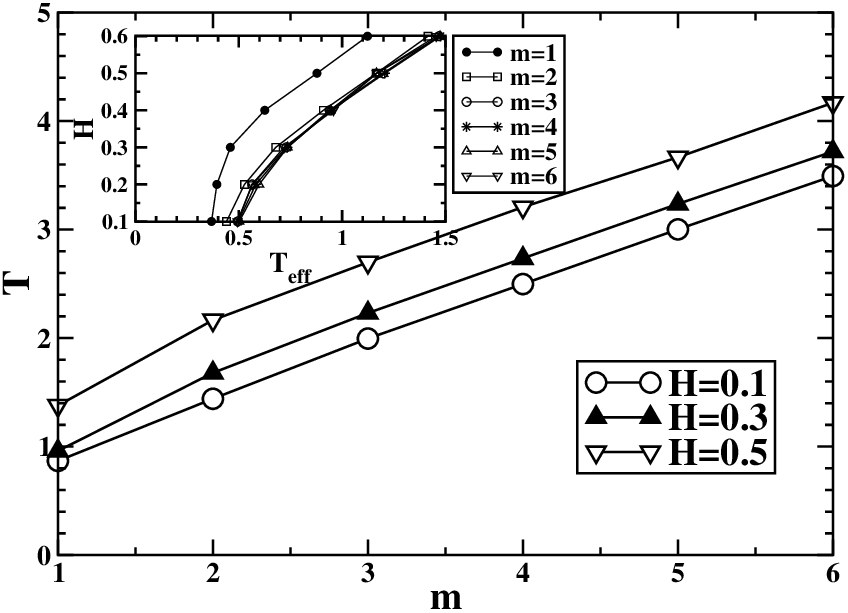}}}
\caption{Order-disorder transition temperature as a function of the Barab\'asi-Albert parameter $m$ for different values of 
the external field, as indicated. The inset shows the phase diagram $H$ vs $T_{eff}$, where the effective 
temperature $T_{eff}= T-0.50\times m$ leads to an approximate data collapse for scale-free networks with $m>1$.} 
\end{figure}
 
Figure 9 shows the transition temperature as a function of the BA parameter $m$ for different values of 
the external field, as indicated. Except for the $m=1$ case, the monotonic shift towards higher temperatures 
appears to be roughly independent of the external field, which suggests to define an effective, $m$-induced 
temperature shift $\Delta T_{eff}(m)=-\alpha m$. 
Using $\alpha=0.5$, we obtain an approximate data collapse for scale-free networks with $m>1$, as shown in the inset 
to Figure 8. Comparing to the small-world network data shown in Figure 4, we confirm that, from a qualitative point of view, 
the critical behavior induced by random connections is generally characteristic of complex network substrates.   

\begin{figure}[t!]
\center \includegraphics[width=3.5truein, height=4.7truein]{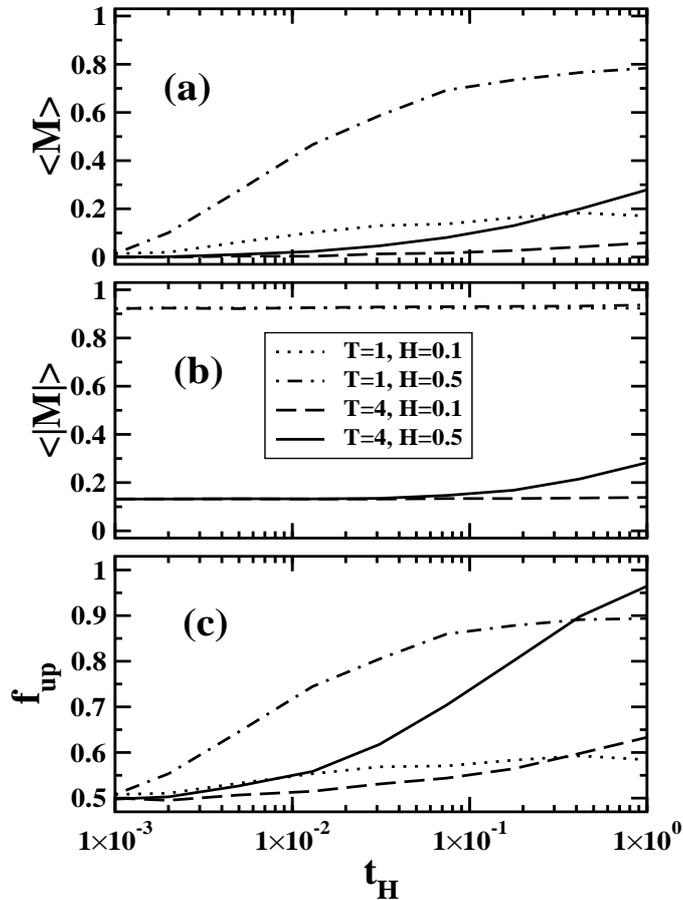}
\caption{Dependence of the ensemble-averaged final states ($t=1$) on the field switch-off time, $t_H$, for different 
temperatures, external fields and observables: (a) the total magnetization, 
(b) the absolute magnetization, and (c) the fraction of final states with a majority of up spins. These results are 
for scale-free networks with BA parameter $m=3$.}
\end{figure}

Figure 10 shows the dependence of the ensemble-averaged final states as a function of  
the field switch-off time, $t_H$, for different temperatures\footnote{Since, for BA-SF networks with $N=10^3$ and $m=3$, 
the zero-field transition is located at $T_c\approx 2$ \cite{can07b}, 
the chosen temperatures roughly correspond to $T=0.5T_c$ and $T=2T_c$, respectively.}, 
external fields and observables: (a) the total magnetization, 
(b) the absolute magnetization, and (c) the fraction of final states with a majority of up spins. The results shown 
correspond to scale-free networks with BA parameter $m=3$. From a qualitative point of view, these results compare well 
to the small-world data shown above (see Figure 6 and discussion).  

\begin{figure}[t]
\centerline{{\epsfxsize=3.9in \epsfysize=2.6in \epsfbox{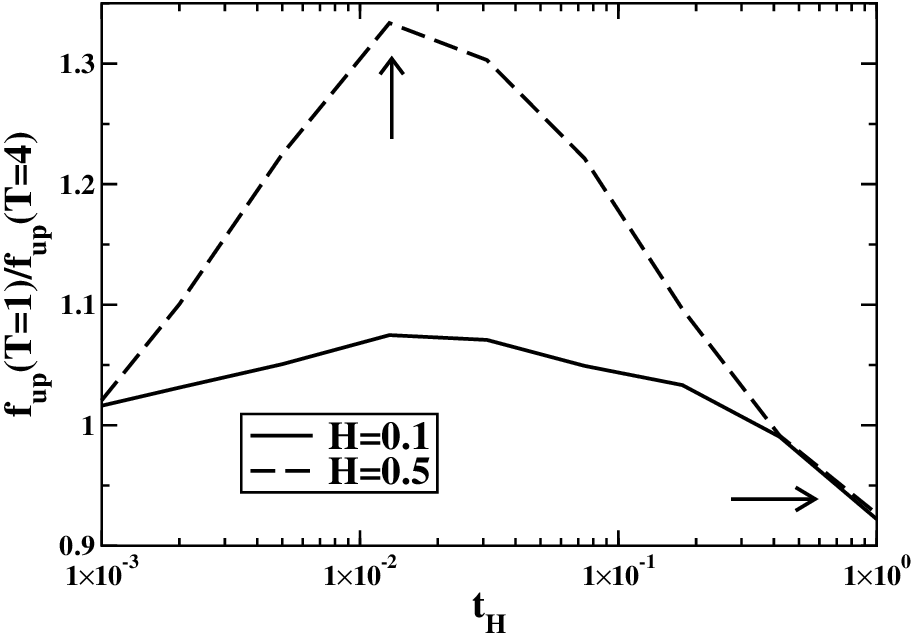}}}
\caption{Ratio $f_{up}(T=1)/f_{up}(T=4)$ as a function of the field switch-off time $t_H$ for scale-free networks with 
BA parameter $m=3$ under different external fields, as indicated.}
\end{figure}

Finally, let us consider in Figure 11 the ratio $f_{up}(T=1)/f_{up}(T=4)$ as a function of $t_H$ for different fields, 
focusing again on scale-free networks with BA parameter $m=3$. 
For $H=0.1$, the ratio shows a slight increase for small values of $t_H$ up to about $t_H\simeq 3\times10^{-2}$, 
while it decreases monotonically for higher values of $t_H$ and reaches a minimum at $t_H=1$ (horizontal arrow). 
This behavior resembles the observations made above for SWNs (recall Figure 7), although the effect apparent here is milder. 
For $H=0.5$, the ratio peaks at $t_H\approx 1.5\times10^{-2}$ (vertical arrow) and then decreases for larger values of $t_H$. 
This confirms that, irrespective of the topological details of the complex network substrate, short, intense campaigns 
are the best strategy to successfully convey a message across a highly cohesive society. 
Incidentally, notice also that, unlike the SWN case, the $t_H\to 1$ tail seems to be roughly independent of the external field.   

\section{Conclusions and Outlook}
In order to investigate advertising and irreversible opinion spreading phenomena, 
we studied the magnetic Eden model (MEM) growing on small-world and scale-free networks with externally applied magnetic fields.  
In this nonequilibrium model, the opinion or decision of an individual 
is affected by those of their acquaintances, but opinion changes (analogous to spin flips in an Ising-like model) 
do not occur. 

The model studied in this work could be realistically applied to sociological scenarios in which individuals are subject 
to highly polarized, short term, binary choice situations under the influence of advertising/marketing mass media campaigns. 
Given these conditions, opinions are not expected to fluctuate and ``thermalize". Some examples are binary voting situations, 
such as a ballotage or referendum, and market competition scenarios between two leading products, where the
validity of this model could be empirically tested using polls, surveys and/or sales performance data.
 
The interplay and competition between the intrinsic local disorder and the externally applied ordering field was 
investigated by means of order-parameter probability distributions and ensemble averages, leading to a temperature 
vs. magnetic field order-disorder phase diagram. The small-world network's shortcut fraction was found to shift the 
effective temperature of the system, suggesting a simple scaling relation that allowed a collapse of all small-world 
network data.  Later, the effects of advertising campaigns with variable duration was discussed. 
The best cost-effective strategy to achieve a desired majority consensus was found to depend crucially on the intrinsic cohesion 
tendency of the society. In the scenario of strongly cohesive social groups, intense, short campaigns are observed to 
be more successful than mild, long ones. However, the opposite conclusion holds in the case of weakly cohesive societies.

The observed phenomena, first presented and discussed for small-world network substrates, were compared to 
corresponding results obtained using scale-free network geometries. We found that, 
irrespective of the details of the underlying complex network topology, consensus and order can be reached for any level of 
intrinsic cohesion and interconnectedness. Moreover, the effects of advertising campaigns with variable duration, 
as well as our conclusions regarding optimal strategies to achieve a majority consensus, remain valid irrespective 
of the specific type of complex network substrate. 

The magnetic Eden model, being different in nature from related equilibrium spin systems such as the Ising model, 
can certainly provide valuable, complementary insight into dynamical and critical aspects of the spreading of opinions 
in a society. Since a characteristic feature of the magnetic Eden model is the absence of thermalization, it may be 
interesting to study a hybrid model with an external parameter controlling the rate of thermalization relative 
to the rate of growth. This generalized model could be tuned from pure MEM-like growth to pure Ising-like thermalization, 
which might be an appropriate representation of sociological scenarios where the opinion of individuals is neither 
completely frozen nor fully thermalized. 

Hopefully, the present findings will thus contribute to the growing interdisciplinary efforts in 
the fields of sociophysics, complex networks, and nonequilibrium statistical physics, and stimulate further work.

\section*{Acknowledgments} 
This work was supported by the James S. McDonnell Foundation and the National 
Science Foundation ITR DMR-0426737 and CNS-0540348 within the DDDAS program, as well as by 
CONICET, ANPCyT, and UNLP (Argentina).

\end{document}